\documentclass[10pt]{article}
\usepackage[utf8]{inputenc}
\usepackage[T1]{fontenc}
\usepackage{authblk}
\usepackage[a4paper, total={6in, 8in}]{geometry}
\usepackage[font=small,labelfont=bf]{caption}
\usepackage{hyperref}
\usepackage{appendix}
\usepackage{cite}
\usepackage{graphicx}
\usepackage{amsmath}
\usepackage{amssymb}
\usepackage{arydshln}
\usepackage{booktabs}
\usepackage{subcaption}
\usepackage{footnote}
\usepackage{fancyhdr}
\usepackage{xcolor}
\usepackage{indentfirst}

\title{Fermion Zero Modes and Fermion number $1/2$ of the 't~Hooft--Polyakov Monopole}
\author[]{P.~E.~Mogaddam\thanks{p\_eslambolchi@sbu.ac.ir} } 

\author[]{S.~S.~Gousheh\thanks{ss-gousheh@sbu.ac.ir}}

\affil[]{\textit{Department of Physics, Shahid Beheshti University, Tehran, Iran}}

\date{\today}

\begin{document}		
	\maketitle
\begin{abstract}
Fermion bound states in the background of the 't~Hooft--Polyakov SU(2) monopole are investigated for various values of gauge coupling constant $g$, the Higgs self-coupling constant $\lambda$, and the Yukawa coupling constant $y_q$. Numerical solutions to the set of coupled differential equations for various selected points in the parameter space reveal only a zero mode, for which we also present an analytic argument. We show that the monopole profile functions and the zero mode wave function become more localized with increasing $g$ and $\lambda$, while the right-handed component of the latter decreases with $g$. However, as expected, this component increases with $y_q$.
We find that the zero mode in the limit $g\to 0$ differs from the zero mode held by the Higgs alone, highlighting the nonlinear and nonperturbative character of the system. Finally, we prove the spectral mirror symmetry of the fermion, whence, together with the existence of the zero mode, we infer the fermion number $1/2$ of the 't~Hooft--Polyakov monopole.  \\
\\
\noindent\textbf{Keywords:}
't~Hooft--Polyakov monopole, fermion zero modes, half-integer fermion number, spectral mirror symmetry.
\end{abstract}

\section{Introduction}
The notion of monopoles can be traced back to Dirac \cite{Dirac:1931kp}, who introduced the concept of a magnetic charge within the framework of U(1) electrodynamics. As noted by Dirac and later reformulated by Wu and Yang \cite{WU1976365}, a magnetic monopole at the origin can be described by two locally regular potentials $\mathbf{A^\pm}$. The potential $\mathbf{A^+}(\mathbf{A^-})$ is regular everywhere except on the negative(positive) $z-$axis---the so-called Dirac string singularity. The Abelian monopole was then generalized to the non-Abelian gauge theories by Wu and Yang \cite{WuYang:1969}, who constructed a monopole solution in the pure SU(2) Yang-Mills theory, where the Dirac string can be globally removed, but the gauge potential remains singular at the origin. Subsequently, 't Hooft \cite{tHooft:1974kcl} and Polyakov \cite{Polyakov:1974ek} independently discovered that introducing a Higgs field in the adjoint representation eliminates this singularity, leading to a regular, finite-energy monopole. Such an SU(2) Yang-Mills-Higgs theory, originally proposed by Georgi and Glashow \cite{PhysRevLett.28.1494}, is a prototype of the electroweak (EW) theory of Weinberg \cite{Weinberg:1967tq} and Salam \cite{Salam1968} with a complex doublet Higgs field. However, it had long been believed that the latter admits no topological monopole of physical relevance, until a non-Abelian, spherically symmetric monopole solution in the full EW theory was discovered by Cho and Maison \cite{ChoMaison1997}.  This solution is similar to the 't~Hooft--Polyakov monopole, but involves a Higgs field in the fundamental representation rather than in the adjoint representation \cite{Gervalle_2022}.

These monopoles can modify the fermion spectrum and even support zero-energy bound states, as has been demonstrated in a variety of other solitonic backgrounds. The study of fermion zero modes in non-trivial backgrounds has a rich history, beginning with the seminal work of Jackiw and Rebbi \cite{JackiwRebbi1976}, who demonstrated that fermions in topological soliton backgrounds can exhibit isolated, normalizable zero-energy modes, leading to fractional fermion number. Jackiw and Rossi \cite{JackiwRossi1981} extended these ideas to vortices, showing that zero modes also appear in lower-dimensional soliton backgrounds and connecting their existence to the index-theorem \cite{Atiyah:1963zz} arguments. Some exactly solvable $(1+1)$ dimensional fermion--soliton systems have also been studied, analyzing their vacuum polarization, spectral structure, and, specifically, their zero modes (see, for example, \cite{PhysRevD.30.2194, gousheh1994vacuum,Shahkarami_2011,charmchi2014complete,charmchi2014massive}). Related works further investigated fermion zero modes on a broad class of solitonic configurations, including domain walls and electroweak strings \cite{Stojkovic_2000, Starkman_2001, Stojkovic_2001, Starkman_2002}, BPS and Kaluza-Klein monopoles \cite{KKMonopole2018}, as well as lattice and supersymmetric settings, demonstrating that the topological protection of zero modes is robust across diverse gauge and Higgs field configurations. More recently, experimental realizations of monopole-bound zero modes in Dirac and acoustic crystals \cite{MonopoleTopoMode2024} have provided a tangible analogue of the original Jackiw-Rebbi predictions, highlighting the relevance of these topological states in both high-energy and condensed matter physics.

In the context of monopole backgrounds, the existence of fermion zero modes in the background of the 't~Hooft--Polyakov monopole has been established only in special limits, such as the Prasad-Sommerfield limit $(\lambda=0)$, for which the corresponding Dirac equations become exactly solvable~\cite{Callias1977}. Despite the significance of this monopole as the prototype of its EW counterpart, a comprehensive investigation of these zero modes beyond the known analytical limits appears to be lacking in the literature. In this work, we systematically analyze the fermion zero modes in the 't~Hooft--Polyakov monopole background, considering various values of the Higgs self-coupling $\lambda$, the gauge coupling constant $g$, and the Yukawa coupling constant $y_q$. We numerically explore how these parameters influence the left-right composition and spatial profiles of the zero modes. Our results reveal that the contribution of the right-handed component to the zero mode starts at zero for $y_q=0$ and increases monotonically with $y_q$. Moreover, by increasing $g$ we obtain a suppression in the right-handed component of the fermion zero mode. We then find exact solution for the zero-energy bound state in the background field of the Higgs alone, and compare it to the limit $g\to 0$ of the zero mode in the monopole background. Finally, we show explicitly that the fermion energy spectrum has mirror symmetry in the presence of a monopole background. To this end, we construct an explicit transformation operator, motivated by Ref.~\cite{Mehraeen_2020}, combining $\gamma^{5}$ with the CP transformation. Under this transformation, the Dirac Hamiltonian is odd while the fermion zero mode is invariant. These are sufficient conditions, according to Jackiw and Rebbi \cite{JackiwRebbi1976}, to infer the fermion number one half for the monopole.

The remainder of this paper is organized as follows. Section ~\ref{monopole} is dedicated to introducing the model, including the equations of motion and the configurations of the monopole and fermions. We then solve the fermion equations numerically, and obtain for each value of $g$, $\lambda$ and $y_q$ exactly one bound state whose energy is zero, a result which we also verify analytically. In Sec.~\ref{paraspace}, we show the dependence of the zero mode wave function on various parameters. In particular, we present a comparison between the contributions of the right- and left-handed components of the zero modes for different values of $g$, $\lambda$ and $y_q$.
In Sec.~\ref{fermion_number}, we construct a transformation that manifests the spectral mirror symmetry and infer the half-integer fermion number for the 't~Hooft--Polyakov monopole. The proof of the spectral mirror symmetry of the fermion spectrum is presented in App.~\ref{appp}. Finally, our conclusions are given in Sec.~\ref{conclusion}.

\section{The Model}\label{monopole}
We consider an extension of the $SU(2)_L$ Yang--Mills--Higgs theory, with the Higgs field being in the adjoint representation, that includes a fermionic sector composed of the up and down quarks with $y_u=y_d=y_q$. The model can be described by the following Lagrangian:
	\begin{align}
		\mathcal{L}=&-\frac{1}{4}F^{\mu \nu}_a F_{\mu \nu}^{a} +D_{\mu} \Phi^{\dagger} D^{\mu}\Phi-\lambda\left(\Phi^{\dagger} \Phi-\frac{v^{2}}{2} \right)^{2}\notag\\&+Q^\dagger_L\, i \sigma^{\mu} D_{\mu}Q_L+Q^\dagger_R \, i \bar{\sigma}^{\mu}\partial_\mu Q_R - y_q Q^\dagger_R \Phi\, Q_L- y_q Q^\dagger_L\Phi^\dagger Q_R, \label{L}  
	\end{align}
	 with the covariant derivatives,
	 \begin{align*}
	 	D_{\mu}Q_L=\partial_{\mu} Q_L+igA_{\mu} Q_L,\qquad D_{\mu}\Phi=\partial_{\mu} \Phi+ig[A_{\mu},\Phi],	
 	\end{align*}
	and the gauge field strength tensor,
	\begin{align}
		F_{\mu\nu}^{a}=\partial_{\mu}A_{\nu}^{a}-\partial_{\nu}A_{\mu}^{a}
		-g\,\varepsilon^{abc}A_{\mu}^{b}A_{\nu}^{c}, \label{F}
	\end{align}
 where $Q_L=(u_L\,,\,d_L)^T,\, Q_R=(u_R\,,\,d_R)^T,\, A_{\mu}=A_{\mu}^a\tau^a/2,\, \Phi=\Phi^a\tau^a/2$, and $\{\tau^{a}\}$ denote the isospin Pauli matrices in the fundamental representation, while in the adjoint representation they are given by $(\tau^{b})^{ac} =2i\varepsilon^{abc}$. 
	
Our aim is to investigate the effects of a fixed standard monopole background field on the fermions. That is, we use the approximation that fermion back-reactions on the gauge and Higgs sectors are negligible, an assumption well justified for classical solitons (see, for example, \cite{Shahkarami_2011}). We also consider the limit of vanishing mixing angle. With these approximations, the field equations are determined from the Lagrangian
in Eq.~\eqref{L} as 
\begin{subequations}
\begin{align}
&D_{\nu}^{ac} F^{\nu\mu}_c =-i\frac{g}{2} \left[\Phi^\dagger\tau^a D^\mu\Phi-(D^\mu\Phi)^\dagger\tau^a\Phi\right],\label{aeom}\\&
D^\mu(D_{\mu}\Phi^{a}) =-2 \lambda \left(\Phi^{\dagger} \Phi-\frac{v^2}{2}\right)\Phi^{a},\label{phieom}   \\&i D_{0} Q_{L}+i \sigma^{i} D_{i} Q_{L}=y_q\Phi\,Q_R, \label{qLeom}\\
&i \partial_{0} Q_{R} - i \sigma^{i} \partial_{i} Q_{R}=y_{q}\Phi^\dagger Q_L \label{Reom}
\end{align} \label{eom}
\end{subequations}

The 't~Hooft--Polyakov monopole \cite{tHooft:1974kcl,Polyakov:1974ek} is a stable, static, finite-energy, and topologically nontrivial solution to the classical equations of motion presented in Eqs.~\eqref{aeom} and \eqref{phieom}.  
In the temporal gauge, where $A_{0}^{a}(x)=0$, this solution is constructed by introducing the following ansatzes:
\begin{subequations}
   \begin{align}
 A_i^a(x)&=\frac{1}{g}\frac{k(r)-1}{r^{2}} \varepsilon_{i j a} x^{j},\label{aanstz}\\
\Phi^a(x)&=i v \sqrt{2}\frac{h(r)}{r} x^a.
\label{hanstz}
\end{align} \label{monoansatz} 
\end{subequations}
Substituting these ansatzes into Eqs.~\eqref{aeom} and \eqref{phieom}, we obtain the following set of coupled nonlinear second order differential equations \cite{manton2004topological,rajaraman1982solitons, Arunasalam_2017}
\begin{subequations}
\begin{align}
&\tilde{k}^{\prime \prime}=\frac{\tilde{k}}{\tilde{r}^{2}}\left(\tilde{k}^2-1\right)+\frac{g^{2}}{2} \tilde{h}^{2}\tilde{k}, \label{keom}\\
&\tilde{h}^{\prime \prime}+\frac{2}{\tilde{r}} \tilde{h}^{\prime}=\frac{2}{\tilde{r}^{2}}\tilde{k}^{2}\tilde{h}+ \lambda\left(\tilde{h}^{2}-1\right)\tilde{h}, \label{heom}
\end{align}
\label{hAeom}
\end{subequations} 
where $\tilde{r}=v r$, $\tilde{A}_\mu(\tilde{r})=A_\mu(r)/v$, ${\tilde\Phi(\tilde{r})=\Phi(r)/v}$, $\tilde{k}(\tilde{r})=k(r)$ and $\tilde{h}(\tilde{r})=h(r)$, with the prime now denoting differentiation with respect to $\tilde{r}$.

Solving Eq.~\eqref{hAeom} requires appropriate boundary conditions for $\tilde{k}$ and $\tilde{h}$. To extract the required boundary conditions, we consider the leading-order polynomial approximation of the profile functions,
 	\[\tilde{k}=c_1 \tilde{r}^n, \qquad \tilde{h}=c_2 \tilde{r}^l,\] 
	which, in the limit $\tilde{r}\to0$, reduce Eq.~\eqref{hAeom} to 
 \begin{subequations}
 		\begin{align}
 	 & n(n-1)=c_1^2\tilde{r}^{2n}-1,\label{k0}\\
 	& l(l-1)+2l=2c_1^2\tilde{r}^{2n}.\label{h0}
 		\end{align}
 \end{subequations}
 For $n<0$, the function $\tilde{k}$ is singular at the origin, and for $n>0$, Eq.~\eqref{k0} turns into $n^2-n+1=0$ which does not have any real roots. Since the gauge field must be real, the only acceptable value of $n$ is zero, leading to $c_1=\pm 1$. Using these values, Eq.~\eqref{h0}  becomes $l^2+l-2=0$, which, besides the unacceptable solution $l=-2$, gives $l=1$. Hence, the boundary conditions at the origin should be $\tilde{k}(0)=1$ \footnote{The case $\tilde{k}(0)=-1$ can be ruled out due to Eq.~\eqref{aanstz}.} and $\tilde{h}(0)=0$. Moreover, the function $\tilde{h}$ should tend to 1 as $\tilde{r}\to\infty$, so that asymptotically $|\Phi|=v/\sqrt{2}$ and the solution has finite energy. In this limit, Eq.~\eqref{keom} becomes $\tilde{k}''=g^2\,\tilde{k}$, whose normalizable solution is $\tilde{k}\sim \exp (-g\,\tilde{r})$, yielding the remaining boundary condition as $\tilde{k}(\infty)=0$.

\begin{figure}[t]
	\begin{subfigure}{.49\textwidth}
		\centering
		\includegraphics[width=1\linewidth]{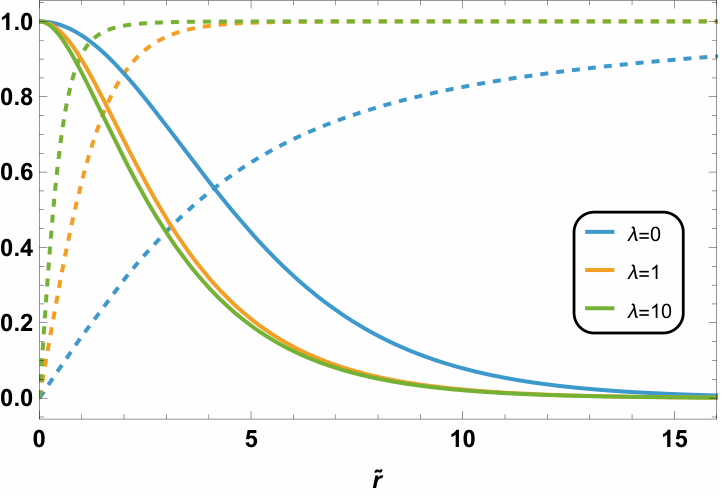}
		\caption{} \label{monofig}
	\end{subfigure}\hfill
 \begin{subfigure}{.49\textwidth}
	\centering
	\includegraphics[width=1\linewidth]{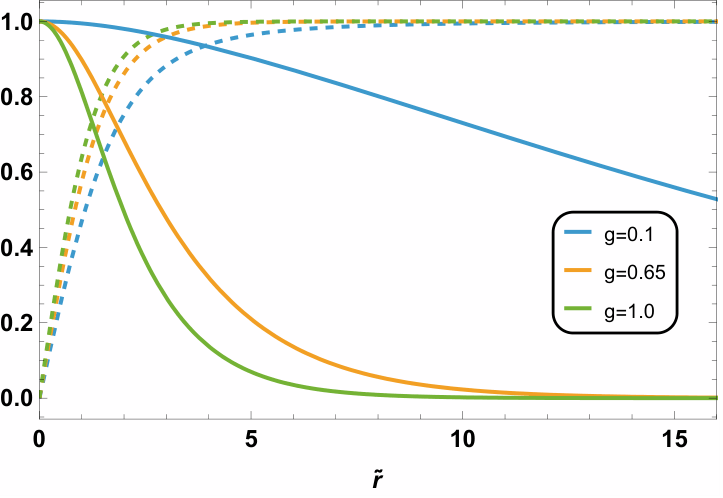}
	\caption{}\label{monog}
\end{subfigure}
\caption{Monopole profile functions, $\tilde{k}$ (solid line) and $\tilde{h}$ (dashed line), in terms of $\tilde{r}$ for (a) the Higgs self-couplings $\lambda=0$ (in blue), 1 (in orange) and 10 (in green), with  $g=0.65$; as well as for (b) the gauge couplings $g=0.1$ (in blue), 0.65 (in orange) and 1.0 (in green), with $\lambda=1$. The upper/lower solid lines and the lower/upper dashed lines are in blue/green.}
\end{figure}

Figure~\ref{monofig} depicts the numerical solutions\footnote{We have employed part of the publicly available Mathematica code shared in Ref.~\cite{MathematicaSE127997}.} of Eq.~\eqref{hAeom} for three different values of the Higgs self-coupling $\lambda$ with $g=0.65$---to be comparable with the approximate value of the EW coupling constant. The monopole functions are also shown in Fig.~\ref{monog} for three different values of $g$ with $\lambda=1$. Comparison of Figs.~\ref{monofig} and \ref{monog} reveals that variations in the Higgs self-coupling, predominantly influence the Higgs field profile, whereas the gauge field profile is governed mainly by the gauge coupling.

Next, we use the solutions of Eqs.~\eqref{aeom} and \eqref{phieom}, examples of which are shown in Figs.~\ref{monofig} and \ref{monog}, as the Higgs and gauge background fields in Eqs.~\eqref{qLeom} and \eqref{Reom}, and adopt the following ansatzes for the fermion fields: 
\begin{align}
		Q_{L/R}=e^{-i \varepsilon t}\left[G_{L/R}(r)+i\boldsymbol{\sigma} \cdot \mathbf{\hat{x}} F_{L/R}(r)\right] \chi_{h}, \label{fermiansatz}
\end{align}
with
\begin{align*}
 \chi_h=\frac{1}{\sqrt{2}}\left[\binom{1}{0}_{S}\binom{0}{1}_{I}-\binom{0}{1}_{S}\binom{1}{0}_{I}\right],
\end{align*}
where $S$ and $I$ refer to spin and isospin, respectively. Here, the spinor $\chi_{h}$ has the hedgehog spin-isospin structure, satisfying the relation
\begin{align}
    (\boldsymbol{\tau}+\boldsymbol{\sigma})\,\chi_{h}=0, \label{grandspin}
\end{align}
which facilitates the reduction of the field equations to their radial forms. 

We utilize the fermion ansatzes introduced in Eq.~\eqref{fermiansatz} together with the monopole configuration given by Eq.~\eqref{monoansatz} and simplify Eqs.~\eqref{qLeom} and \eqref{Reom}, obtaining,
\begin{align}
	& {\left[\begin{array}{cccc}
			0 & \frac{d}{d \tilde{r}}+\frac{1+\tilde{k}}{\tilde{r}} & 0 & \frac{y_q}{\sqrt{2}} \tilde{h} \\
			-\frac{d}{d \tilde{r}}-\frac{1-\tilde{k}}{\tilde{r}}& 0 & -\frac{y_q}{\sqrt{2}} \tilde{h} & 0 \\
			0 & -\frac{y_q}{\sqrt{2}} \tilde{h} & 0 & -\frac{d}{d \tilde{r}}-\frac{2}{\tilde{r}} \\ \frac{y_q}{\sqrt{2}} \tilde{h} & 0 & \frac{d}{d \tilde{r}} & 0
		\end{array}\right]\left(\begin{array}{c}
			\tilde{G}_{L} \\
			\tilde{F}_{L} \\
			\tilde{G}_{R} \\
			\tilde{F}_{R}
		\end{array}\right)=\tilde{\varepsilon}\left(\begin{array}{c}
			\tilde{G}_{L} \\
			\tilde{F}_{L} \\
			\tilde{G}_{R} \\
			\tilde{F}_{R}
		\end{array}\right)}, \label{equmat}
\end{align}
with the eigenvalue $\tilde{\varepsilon}=\varepsilon/v$, and $\tilde{G}(\tilde{r})=G(r)/v^{3/2}$ and $\tilde{F}(\tilde{r})=F(r)/v^{3/2}$.

To extract the appropriate boundary conditions, we analyze the behavior of the equations in the limit $\tilde r \to 0$, while requiring all wave functions
$\tilde{G}_{L}$, $\tilde{F}_{L}$, $\tilde{G}_{R}$, and $\tilde{F}_{R}$ to vanish as $\tilde r \to \infty$ in order to have normalizable bound states. Near the origin, Eq.~\eqref{equmat} reduces to 
 \begin{subequations}
 \begin{align}
  &\tilde{G}_{L/R}'=\mp\tilde{\varepsilon}\,\tilde{F}_{L/R},\label{originG} \\
  &\tilde{F}_{L/R}'+\frac{2}{\tilde{r}}\tilde{F}_{L/R}=\pm\tilde{\varepsilon}\,\tilde{G}_{L/R},\label{originF}
 \end{align}\label{originGF}
  \end{subequations}
where we have used $\tilde{k}\to 1$ and $\tilde{h}\to 0$ as $\tilde{r}\to0$.
The system of equations~\eqref{originGF} can now be decoupled and expressed as a set of second-order differential equations:
\begin{align}
  \tilde{G}''_{L/R}
    +\frac{2}{\tilde{r}} \tilde{G}_{L/R}'=-\tilde{\varepsilon}^2\,\tilde{G}_{L/R}, \qquad \tilde{F}''_{L/R}
    +\frac{2}{\tilde{r}} \tilde{F}_{L/R}'-\frac{2}{\tilde{r}^2}\tilde{F}_{L/R}=-\tilde{\varepsilon}^2\,\tilde{F}_{L/R}.\label{originGL2}
\end{align}
Considering the leading-order polynomial approximation to the wave functions,
\[\tilde{G}=c_1\, \tilde{r}^n,\qquad \tilde{F}=c_2\, \tilde{r}^l,\]
Eq.~\eqref{originGL2} turns into the equations
$n(n+1)=0$ and $l(l+1)-2=0$, yielding $n=0,-1$ and $l=1,-2$. Therefore, the acceptable solutions must have the following behavior as $\tilde r \to 0$\textcolor{blue}{:}\textcolor{red}{\,,} \[\tilde{G}=c_1, \qquad \tilde{F}=c_2 \tilde{r}.\]

Inserting the above solutions into the original first-order system, Eq.~\eqref{originF} yields
$3c_2 =\pm \tilde{\varepsilon} c_1$, while
Eq.~\eqref{originG} gives $\tilde{\varepsilon} c_2 \tilde r = 0$, which implies $ \tilde{\varepsilon}c_2= 0$. Thus, the nontrivial solution must have $c_2=0$, $\tilde{\varepsilon} = 0$, and $c_1 \ne 0$\footnote{The conditions $\tilde{F}_{L/R}(0)=0$ are in fact expected, since the functions $\tilde{F}_{L/R}$ multiply the spinor structure  $\boldsymbol{\sigma}\cdot\hat{\mathbf{x}}$ in the fermion ansatz, and these conditions eliminate the ambiguity associated with defining $\hat{\mathbf{x}}$ at the origin.}.
Such a solution with $\tilde{G}_{L/R} \neq 0$ exists only at zero energy and is necessarily accompanied by
$\tilde{F}_{L/R} = 0$, not as an additional assumption, but as a direct consequence of imposing regularity at the origin. This conclusion is fully supported by our numerical analysis, which shows that for arbitrary Yukawa and Higgs self-couplings the spectrum contains a single bound state, corresponding to the zero mode. It thus suffices to compute the eigenvector of Eq.~\eqref{equmat} associated with $\tilde{\varepsilon}=0$, whose irreducible form can be written as
\begin{align*}
	\left(\begin{array}{cc}
		\frac{d}{d \tilde{r}}+\frac{1+\tilde{k}}{\tilde{r}} & 0 \\
		0 & \frac{d}{d \tilde{r}}+\frac{1-\tilde{k}}{\tilde{r}}
	\end{array}\right)\binom{\tilde{F}_{L}}{\tilde{G}_{L}}&=-\frac{y_q}{\sqrt{2}} \tilde{h}\binom{\tilde{F}_{R}}{\tilde{G}_{R}},\notag \\ \left(\begin{array}{cc}
		\frac{d}{d \tilde{r}}+\frac{2}{\tilde{r}} & 0 \\
		0 & \frac{d}{d \tilde{r}}
	\end{array}\right)\binom{\tilde{F}_{R}}{\tilde{G}_{R}}&=- \frac{y_q}{\sqrt{2}}\tilde{h}\binom{\tilde{F}_{L}}{\tilde{G}_{L}}.
\end{align*}
Our numerical analysis further confirms that the functions $\tilde{F}_{L/R}$ for all fermion zero modes vanish identically. Moreover, as the $\tilde{F}_{L/R}$ components are responsible for generating the angular momentum, their absence restricts the system to $\mathbf{L}=0$, thereby extending Eq.~\eqref{grandspin} to the familiar grand–spin relation $\mathbf{T}+\mathbf{S}+\mathbf{L}=0$ for the zero modes in topologically nontrivial backgrounds, 
with $\mathbf{T}=\boldsymbol{\tau}/2$ and $\mathbf{S}=\boldsymbol{\sigma}/2$ \cite{JackiwRebbi1976,PhysRevD.26.2058, PhysRevD.45.2920}. Taking these into account, our field equations reduce to,
\begin{align}
	\left(\begin{array}{cc}
		\frac{d}{d \tilde{r}}+\frac{1-\tilde{k}}{\tilde{r}} & \frac{y_q}{\sqrt{2}} \tilde{h} \\
		\frac{y_q}{\sqrt{2}} \tilde{h} & \frac{d}{d \tilde{r}}
	\end{array}\right)\binom{\tilde{G}_{L}}{\tilde{G}_{R}}&=\binom{0}{0}. \label{firstorderreduced}
\end{align}
At this stage, one can define a mass parameter for the quarks as $m_q=v\tilde{m}_q=v y_q\tilde{h}(\infty)/\sqrt{2}= v y_q/\sqrt{2}$, merely for the purpose of comparing various scales. In the next section, we shall display the numerical solutions to this equation for various values of the parameters $\{g, \lambda, y_q \}$.

\section{Numerical Solutions} \label{paraspace}
In this section, we numerically solve Eq.~\eqref{firstorderreduced} to investigate the effects of the gauge coupling $g$, the Higgs self-coupling $\lambda$, and the Yukawa coupling $y_q$ on the 't~Hooft--Polyakov monopole and its fermion zero mode. Though not necessary, we assume that $y_{q}\ll 1$ so that the mass parameter of the quarks $\tilde{m}_q=y_q/\sqrt{2}\ll 1$. 
We then present our results for some selected points in the parameter space in the range: $0.1\leq g\leq1,\, 0\leq\lambda\leq10,\, 0.01\leq y_q\leq0.1$.

\begin{figure}[t]
\begin{subfigure}{.49\textwidth}
	\centering
\includegraphics[width=1\linewidth]{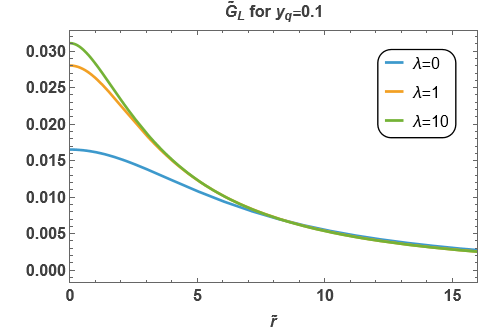}
		\caption{}
        \end{subfigure}\hfill
\begin{subfigure}{.49\textwidth}
	\centering	\includegraphics[width=1\linewidth]{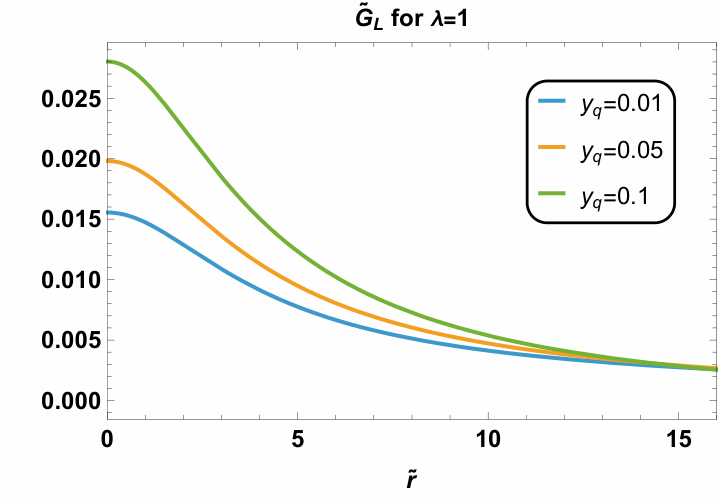}
	\caption{}
    \end{subfigure}

\medskip

\begin{subfigure}{.49\textwidth}
		\centering
\includegraphics[width=1\linewidth]{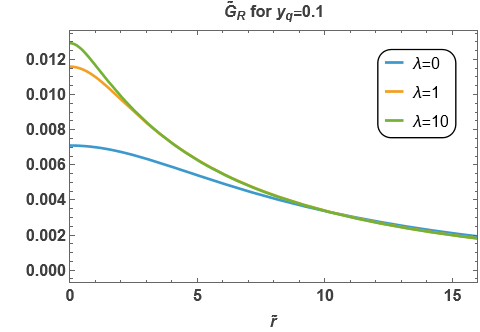}
		\caption{}
	\end{subfigure}\hfill
 \begin{subfigure}{.49\textwidth}
	\centering	\includegraphics[width=1\linewidth]{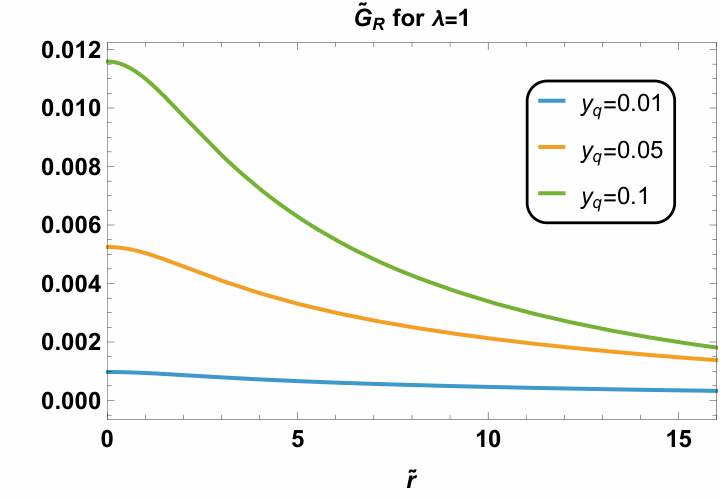}
	\caption{}
\end{subfigure}%
	\caption{The dependence of the left-handed component $\tilde{G}_L$ on (a) the Higgs self-coupling $\lambda$ for fixed $y_q=0.1$, and on (b) the Yukawa coupling $y_q$ for fixed $\lambda=1$; as well as the dependence of the right-handed component $\tilde{G}_R$ on (c) $\lambda$ for fixed $y_q=0.1$, and on (d) $y_q$ for fixed $\lambda=1$, all with $g=0.65$.} \label{GRGLfig} 
\end{figure}

Setting $g=0.65$, Fig.~\ref{GRGLfig} demonstrates the dependence of the $\tilde{G}_{L/R}$ functions on $\lambda$ for fixed $y_q$, as well as on $y_q$ for fixed $\lambda$. 
Our numerical results indicate that a larger $\lambda$, which is related to the Higgs mass parameter by $m_H=v\sqrt{2\lambda}$, or $y_q$ enhances the spatial localization of the fermion zero modes.

\begin{figure}[t]
    \centering    
    \includegraphics[width=0.51\linewidth]{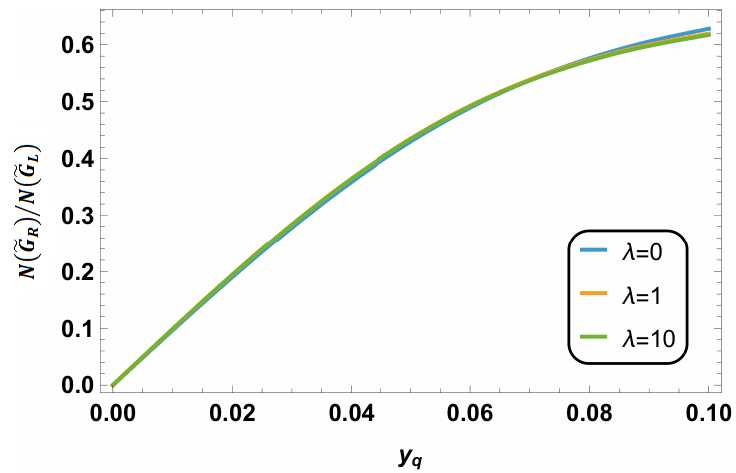}
    \caption{The ratio of the norms of the right- and left-handed components of the fermion zero mode $N(\tilde{G}_R)/N(\tilde{G}_L)$ as a function of $y_q$ for different values of $\lambda$, with $g=0.65$. The right-handed contribution increases with fermion mass parameter, but is essentially independent of $\lambda$.}
    \label{ratio-lambda}
\end{figure}

Figure~\ref{ratio-lambda} displays the relative norms of the right- and left-handed components of the fermion zero mode in terms of $y_q$ for different values of $\lambda$, confirming that the massless zero modes are purely left-handed. From Fig.~\ref{ratio-lambda}, it is also evident that the right-handed components of the fermion zero modes are increased by increasing $y_q$ at fixed $\lambda$, while they are nearly independent of  $\lambda$ at fixed $y_q$.

\begin{figure}[t]
	\begin{subfigure}{.48\textwidth}
		\centering
		\includegraphics[width=1\linewidth]{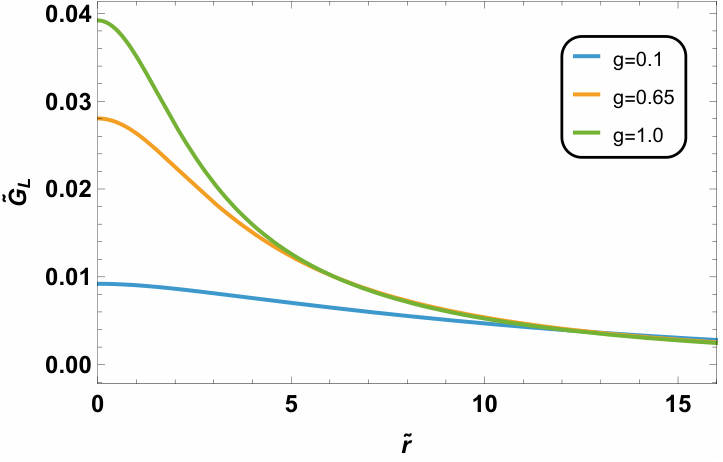}
		\caption{}
	\end{subfigure}\hfill
    \begin{subfigure}{.48\textwidth}
	\centering
\includegraphics[width=1\linewidth]{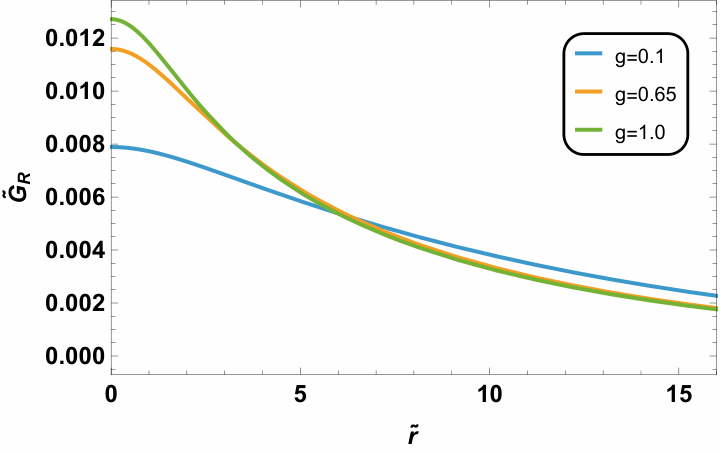}
	\caption{}
\end{subfigure}%
	\caption{Dependence of (a) $\tilde{G}_L$ and (b) $\tilde{G}_R$ on $g$, for $\lambda=1$ and $ y_q=0.1$.} \label{Gg}
\end{figure}

\begin{figure}[t]
	\centering
	\includegraphics[width=0.51\linewidth]{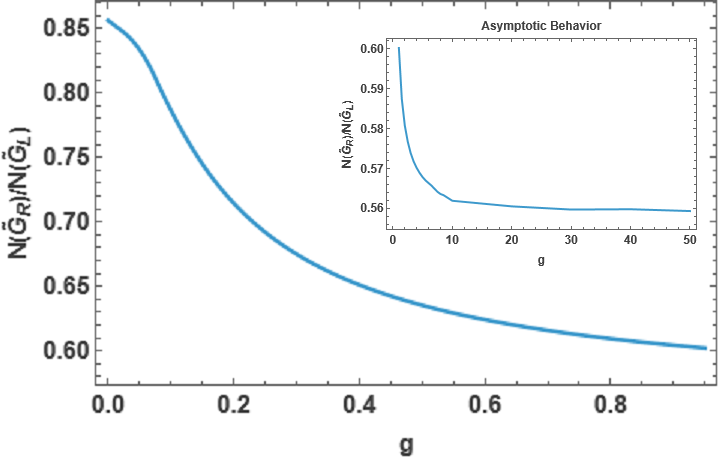}
\caption{The ratio of the norms of the right- and left-handed components of the fermion zero mode $N(\tilde{G}_R)/N(\tilde{G}_L)$ as a function of $g$, for $\lambda=1$ and $y_q=0.1$. The contribution of the right-hand component decreases with increasing the $g$, approaching an asymptotic value. } \label{ratio-g}
\end{figure}

Setting $\lambda=1$ and $y_q=0.1$, the $\tilde{G}_{L/R}$ functions are plotted in Fig.~\ref{Gg} for the three different values of $g$. As can be seen from this figure, the larger the value of $g$ the more localized the wave functions for the zero mode are. The ratio of the norms of $\tilde{G}_R$ to $\tilde{G}_L$ is shown in Fig.~\ref{ratio-g}, which shows that as $g$ increases, the value of $N(\tilde{G}_{R})/N(\tilde{G}_{L})$ reduces and relaxes into a constant.  
On the other hand, as $g$ decreases, the norm of $\tilde{G}_{R}$ grows, while that of $\tilde{G}_{L}$ decreases, such that their ratio approaches another constant at $g=0$. The values of these constants depend strongly on $y_q$ and weakly on $\lambda$.  

One might have expected that the value of the latter constant, {\it i.e.}, $N(\tilde{G}_{R})/N(\tilde{G}_{L})|_{g=0}$, to be equal to $1$, since in the limit $g \to 0$, Fig.~\ref{monog} indicates that the function $\tilde{k} \to 1$ for all finite $\tilde{r}$, and hence the background gauge filed becomes negligible compared to the Higgs field. However, as shown in Fig.~\ref{ratio-g}, this is not the case. To explore this issue, let us first set $\tilde{k} = 1$, which is equivalent to setting $A_{\mu}=0$ from the beginning, for all $\tilde{r}$. Then all of the above arguments leading to $\tilde{F}_{L/R}=0$ and $\tilde{\varepsilon}=0$ still hold true and the system of Eqs.~\eqref{heom} and \eqref{firstorderreduced} reduces to
\begin{subequations}
\begin{align}
	&\tilde{h}''+\frac{2}{\tilde{r}}\tilde{h}'=\frac{2}{\tilde{r}^2}\tilde{h}+\lambda \left(\tilde{h}^2-1\right)\tilde{h}, \label{hnot0}\\
	&\tilde{G}_L'=-\frac{y_q}{\sqrt{2}}\tilde{h}\tilde{G}_R,\qquad \tilde{G}_R'=-\frac{y_q}{\sqrt{2}}\tilde{h}\tilde{G}_L. \label{hnot0f}
\end{align}
\end{subequations}

A nontrivial, normalizable solution to Eq.~\eqref{hnot0f} is 
 \[\tilde{G}_L(\tilde{r})=\tilde{G}_R(\tilde{r})=N \exp \left[-\frac{y_q}{\sqrt{2}}\int_{0}^{\tilde{r}}\mathrm{d}\tilde{r}'\,\tilde{h}(\tilde{r}')\right],\]
in which $\tilde{h}$ satisfies Eq.~\eqref{hnot0}, and the functions $\tilde{G}_{L/R}$ are exponentially damped as $\tilde{r}\to\infty$. Accordingly, if $\tilde{h}\ne0$ and $\tilde{k}=1$, fermion zero modes can still occur in the background of the Higgs field alone, with $\tilde{G}_{R}(\tilde{r})=\tilde{G}_{L}(\tilde{r})$. However, as stated above, in general for the 't~Hooft--Polyakov monopole $N(\tilde{G}_{R})/N(\tilde{G}_{L})|_{g=0}\neq 1$, which indicates that for this nonperturbative and topologically nontrivial configuration the effects of the gauge and Higgs parts cannot be decoupled or disentangled. In fact, as can be seen from Figs.~\ref{monog} and \ref{Gg}, as $g\to 0$, while $\tilde{k}\to 1$ for finite $\tilde{r}$, the bound state wave function spreads out, becoming more sensitive to the large $\tilde{r}$ behavior of the monopole.

\section{Fermion Number 1/2}\label{fermion_number}
In this section, we introduce a transformation operator which manifests the spectral mirror symmetry of the Dirac Hamiltonian in the 't~Hooft--Polyakov monopole background. The Dirac Hamiltonian operator is derivable from $\mathcal{L}$ given by Eq.~\eqref{L} and can be expressed as follows \cite{Klinkhamer_2003}
\begin{align}
	\hat{\mathcal{H}}=-i\gamma^0\gamma^iD_i+y_q\gamma^0\left(\Phi^\dagger P_L+\Phi P_R\right), \label{H}
\end{align} 
where $D_i=\partial_{i}+igA_i^a(\tau^a/2) P_L$, and the projection operators are defined by
\begin{align*}
	P_L=\frac{1-\gamma^5}{2},\qquad 	P_R=\frac{1+\gamma^5}{2},
\end{align*}
with the following choice of Weyl representation for the gamma matrices:
\begin{align*}
	\gamma^0=\begin{pmatrix}
		0 & I_2 \\ I_2 & 0
	\end{pmatrix}, \qquad \gamma^i=\begin{pmatrix}
	0 & \sigma^i \\ -\sigma^i & 0
\end{pmatrix}, \qquad
\gamma^5=\begin{pmatrix}
	-I_2 & 0 \\ 0 & I_2
\end{pmatrix}.
\end{align*}
 
   It is known that the combined transformation of C and P acts non-trivially on the Higgs and fermion fields as
	\begin{align*}
		\Phi(\mathbf{r})\overset{CP}{\longrightarrow}i\tau^2\Phi^*(-\mathbf{r})(-i\tau^2)=\Phi(-\mathbf{r}), \qquad 	\binom{Q_L}{Q_R}(\mathbf{r},t)\overset{CP}{\longrightarrow}i\tau^2\gamma^2\gamma^0 \binom{Q^*_L}{Q^*_R}(\mathbf{-r},t).
\end{align*}
Therefore, the charge-conjugated, parity-inverted Hamiltonian becomes \cite{Mehraeen_2020}
\begin{align}
	\hat{\mathcal{H}}^{CP}(\mathbf{r})=-\gamma^2\gamma^0 \begin{pmatrix}
		-\hat{\mathcal{H}}_{22}^*  & \hat{\mathcal{H}}_{21}^* \\ \hat{\mathcal{H}}_{12}^*  & -\hat{\mathcal{H}}_{11}^*
	\end{pmatrix}_{(-\mathbf{r})} \gamma^0\gamma^2. \label{HCP}
\end{align}

Using the ansatz given in Eq.~\eqref{hanstz}, we construct $\Phi$ and substitute it into Eq.~\eqref{H} to obtain an expression for $\hat{\mathcal{H}}$, as shown in App.~\ref{appp}. Inserting the matrix elements from Eq.~\eqref{Hcomponents} into Eq.~\eqref{HCP}, we find
 \begin{align}
	\hat{\mathcal{H}}^{CP} + \hat{\mathcal{H}}=2y_q\gamma^0\gamma^5\Phi, \label{evenpart} 
\end{align}
	which indicates that $\hat{\mathcal{H}}$ is not odd under CP, except in the trivial case $\Phi= 0$.

 Inspired by Ref.~\cite{Mehraeen_2020}, we now define a new conjugation operator, $\widetilde{CP}$, consisting of the CP transformation along with an additional $\gamma^5$ as 
 \begin{align}
 	\widetilde{CP} \equiv -i\gamma^5 CP. \label{tildeCP}
 \end{align}   
In App.~\ref{appp}, we repeat the calculation leading to Eq.~\eqref{evenpart} for the transformation introduced in Eq.~\eqref{tildeCP}, yielding 
\begin{align}
	\hat{\mathcal{H}}^{\widetilde{CP}}=-\hat{\mathcal{H}}. \label{HCPtilde}
\end{align}
The existence of a transformation under which $\hat{\mathcal{H}}$ is odd ascertains the spectral mirror symmetry. That is, under such a transformation every eigenstate of $\hat{\mathcal{H}}$ with energy $\varepsilon$ is transformed into one with energy $-\varepsilon$, the only exception being a zero energy bound state which must then be invariant under such a transformation. Thus, an important consistency check of our symmetry transformation would be to operate it on the fermion zero mode identified in our analysis, {\it i.e.},
	\begin{align}
	\binom{Q_{0,L}}{Q_{0,R}}(\mathbf{r})\overset{\widetilde{CP}}{\longrightarrow} \tau^2\gamma^5\gamma^2\gamma^0\binom{Q^*_{0,L}}{Q^*_{0,R}}(-\mathbf{r})=-\tau^2\begin{pmatrix}
			\sigma^2 & 0 \\ 0 & \sigma^2
		\end{pmatrix}\binom{Q_{0,L}^*}{Q_{0,R}^*}(-\mathbf{r}). \label{Phi0'}
\end{align}
 Here, $Q_{0,L/R}$ denote the corresponding ansatzes for the fermion zero mode given in Eq.~\eqref{fermiansatz}. Consequently, as $F_{L/R}=0$ for the fermion zero modes, ${Q_{0,L/R}^*}(-\mathbf{r})=Q_{0,L/R}(\mathbf{r})$, and Eq.~\eqref{Phi0'} is reduced to 
\begin{align*}
	\chi_h\overset{\widetilde{CP}}{\longrightarrow}-\tau^2\sigma^2\chi_h=\begin{pmatrix}
		0 & -1 \\ 1 & 0
	\end{pmatrix}_I\begin{pmatrix}
	0 & -1 \\ 1 & 0
\end{pmatrix}_S\frac{1}{\sqrt{2}}\left[\binom{1}{0}_{S}\binom{0}{1}_{I}-\binom{0}{1}_{S}\binom{1}{0}_{I}\right]=\chi_h,
\end{align*}
verifying that the fermion zero mode in the 't~Hooft--Polyakov monopole background is indeed $\widetilde{CP}$-invariant. 

\section{Conclusion}\label{conclusion}
In this paper, we have shown that the only fermion bound state in the 't~Hooft--Polyakov monopole background is the zero mode. We have investigated the behavior of the fermion zero mode by numerically solving the corresponding equations of motion for various values of the Yukawa coupling $y_q$, the Higgs self-coupling $\lambda$, and the gauge coupling $g$.
We have shown that by increasing $g$ or $\lambda$, the monopole profile function $\tilde{k}(\tilde{r})$ and $\tilde{h}(\tilde{r})$, as well as the zero mode wave function, become more localized. Also, as $y_q$ increases, the latter adheres more closely to the profile functions which effectively localizes it further.
The fermion zero mode is composed of both left- and right-handed components, such that the contribution of the right-handed part, as expected, increases with $y_q$ and decreases with $g$, and is largely unaffected by $\lambda$.  Finally, by introducing the symmetry operator $\widetilde{CP}$, constructed from the CP transformation supplemented by $\gamma^{5}$, we have established the spectral mirror symmetry of the Dirac Hamiltonian in the 't~Hooft--Polyakov monopole background. As an important consistency check, we have verified that the fermion zero mode obtained in our analysis remains invariant under this transformation. The coexistence of spectral mirror symmetry and the fermion zero mode then provides the two essential conditions required to infer a half-integer fermion number, specifically $1/2$, for the 't~Hooft--Polyakov monopole, in accordance with the Jackiw and Rebbi reasoning.

 A natural extension of this work would be to incorporate the back-reaction of fermions on the monopole field, which may refine the quantitative features of the solutions and offer deeper insight into their structure. It would also be valuable to compare the fermion zero modes of the 't~Hooft--Polyakov monopole with those of the EW theory, if they exist, thereby providing a clearer understanding of the relation between the two models.
 
\appendix 
\section{Spectral Mirror Symmetry}\label{appp}
In this appendix, we first derive the explicit functional forms of the components appearing in Eq.~\eqref{H}. We then construct the CP-transformed Hamiltonian, which yields Eq.~\eqref{evenpart}. The antisymmetric behavior of $\hat{\mathcal{H}}$ under the $\widetilde{CP}$ operation follows directly from this result.

 As an SU(2)-valued matrix, the Dirac Hamiltonian is
 \begin{align*}
 	\hat{\mathcal{H}}=\begin{pmatrix}
 		\hat{\mathcal{H}}_{11} & \hat{\mathcal{H}}_{12} \\ \hat{\mathcal{H}}_{21} & \hat{\mathcal{H}}_{22}
 	\end{pmatrix},
 \end{align*}
and the Higgs matrix of the 't~Hooft--Polyakov monopole can be expressed as
	\begin{align*}
		\Phi=\frac{v}{\sqrt{2}}h(r)\begin{pmatrix}
			i\hat{x}_3 & i(	\hat{x}_1-i\hat{x}_2) \\
			i(\hat{x}_1+i\hat{x}_2) & -i\hat{x}_3
		\end{pmatrix}=\begin{pmatrix}
		\Phi_{22}^* & \Phi_{12} \\
	-\Phi_{12}^* & \Phi_{22}
	\end{pmatrix}, 
\end{align*}
where $\Phi_{22}^*=-\Phi_{22}$.
Then the components of $\hat{\mathcal{H}}$ take the form 
\begin{subequations}
	\begin{align}
		&\hat{\mathcal{H}}_{11}=-i\gamma^0\gamma^i\partial_{i}+\frac{g}{2}A_i^3\gamma^0\gamma^iP_L+y_q\Phi_{22} \gamma^0 (P_L-P_R),\\
		&\hat{\mathcal{H}}_{12}=\frac{g}{2}(A_i^1-i A_i^2)\gamma^0\gamma^iP_L+y_q \Phi^*_{12} \gamma^0 (P_L-P_R),\\
		&\hat{\mathcal{H}}_{21}=\frac{g}{2}(A_i^1+i A_i^2)\gamma^0\gamma^iP_L-y_q \Phi_{12}\gamma^0  (P_L-P_R),\\
		&\hat{\mathcal{H}}_{22}=-i\gamma^0\gamma^i\partial_{i}-\frac{g}{2}A_i^3\gamma^0\gamma^iP_L-y_q \Phi_{22} \gamma^0 (P_L-P_R). 
	\end{align}\label{Hcomponents}
\end{subequations} 

Utilizing Eqs.~\eqref{HCP} and \eqref{Hcomponents} as well as the relations $\mathcal{\hat{H}}(\mathbf{-r})=-\mathcal{\hat{H}}(\mathbf{r})$ and $\gamma^2{\gamma^i}^*\gamma^2=-\gamma^i$, the components of the CP-transformed Hamiltonian, $\hat{\mathcal{H}}^{CP}$, can also be calculated: 
\begin{subequations}
	\begin{align}
		\hat{\mathcal{H}}^{CP}_{11}&=-\gamma^2\gamma^0\hat{\mathcal{H}}_{22}^*\gamma^0\gamma^2=i\gamma^0\gamma^i\partial_{i}-\frac{g}{2}A_i^3\gamma^0\gamma^iP_L+y_q \Phi_{22} \gamma^0 (P_L-P_R), \\
		\hat{\mathcal{H}}^{CP}_{12}&=\gamma^2\gamma^0\hat{\mathcal{H}}_{21}^*\gamma^0\gamma^2=-\frac{g}{2}(A_i^1-i A_i^2)\gamma^0\gamma^iP_L+y_q \Phi^*_{12}\gamma^0 (P_L-P_R),\\
		\hat{\mathcal{H}}^{CP}_{21}&=\gamma^2\gamma^0\hat{\mathcal{H}}_{12}^*\gamma^0\gamma^2=-\frac{g}{2}(A_i^1+i A_i^2)\gamma^0\gamma^iP_L-y_q \Phi_{12} \gamma^0 (P_L-P_R),\\
		\hat{\mathcal{H}}^{CP}_{22}&=-\gamma^2\gamma^0\hat{\mathcal{H}}_{11}^*\gamma^0\gamma^2=i\gamma^0\gamma^i\partial_{i}+\frac{g}{2}A_i^3\gamma^0\gamma^iP_L-y_q\Phi_{22} \gamma^0 (P_L-P_R),
	\end{align}\label{HCPcomponents}
\end{subequations} 
which, along with Eq.~\eqref{Hcomponents}, leads to  Eq.~\eqref{evenpart}.

Furthermore, sandwiching Eq.~\eqref{HCPcomponents} between two $\gamma^{5}$ matrices and using the identities
\begin{align*}
\gamma^5 P_{R/L}=\pm P_{R/L}\gamma^5, \qquad P_{L/R}\gamma^\mu=\gamma^\mu P_{R/L},
\end{align*}
yields Eq.~\eqref{HCPtilde}, demonstrating that the Hamiltonian is odd under the $\widetilde{CP}$ transformation.

\bibliographystyle{ieeetr}
	\bibliography{bib}
\end{document}